\documentclass[a4paper,twocolumn,11pt,unpublished]{quantumarticle}
\pdfoutput=1
\usepackage[utf8]{inputenc}
\usepackage[english]{babel}
\usepackage[T1]{fontenc}
\usepackage{amsmath}
\usepackage[linkcolor=blue,citecolor=blue,urlcolor=blue]{hyperref}
\usepackage{bm}
\usepackage{tikz}
\usepackage{lipsum}
\usepackage{listings}
\usepackage{xcolor}
\usepackage{algorithmic}
\usepackage{algorithm}

\definecolor{codegreen}{rgb}{0,0.6,0}
\definecolor{codegray}{rgb}{0.5,0.5,0.5}
\definecolor{codepurple}{rgb}{0.58,0,0.82}
\definecolor{backcolour}{rgb}{0.95,0.95,0.92}

\lstdefinestyle{mystyle}{
	backgroundcolor=\color{backcolour},   
	commentstyle=\color{codegreen},
	keywordstyle=\color{magenta},
	numberstyle=\tiny\color{codegray},
	stringstyle=\color{codepurple},
	basicstyle=\ttfamily\footnotesize,
	breakatwhitespace=false,         
	breaklines=true,                 
	captionpos=b,                    
	keepspaces=true,                 
	numbers=left,                    
	numbersep=5pt,                  
	showspaces=false,                
	showstringspaces=false,
	showtabs=false,                  
	tabsize=2
}
\usepackage[numbers,sort&compress]{natbib}

\begin{document}

\title{Graphix: optimizing and simulating measurement-based quantum computation on local-Clifford decorated graph}

\author{Shinichi Sunami}
\affiliation{Clarendon Laboratory, University of Oxford, Oxford OX1 3PU, United Kingdom}
\orcid{0000-0002-0969-9909}
\email{shinichi.sunami@physics.ox.ac.uk}
\author{Masato Fukushima}
\affiliation{Department of Physics, Faculty of Science, The University of Tokyo, 7-3-1 Hongo, Bunkyo-ku, Tokyo 113-0033, Japan}
\affiliation{Fixstars Amplify, 3-1-1 Shibaura, Minato-ku, Tokyo 108-0023, Japan}

\maketitle

\begin{abstract}	
We introduce an open-source software library \emph{Graphix}, which optimizes and simulates measurement-based quantum computation (MBQC).
By combining the measurement calculus with an efficient graph state simulator, Graphix allows the classical preprocessing of Pauli measurements in the measurement patterns, significantly reducing the number of operations required to perform the quantum computation while maintaining determinism.
For a measurement pattern translated from a quantum circuit, this corresponds to the preprocessing of all Clifford gates, and this improvement in the one-way model is important for efficient operations in quantum hardware with limited qubit numbers.
In addition to the direct translation from gate networks, we provide a pattern generation method based on flow-finding algorithms, which automatically generates byproduct correction sequences to ensure determinism.
We further implement optimization strategies for measurement patterns beyond the standardization procedure and provide tensor-network backend for classically simulating the MBQC.
\end{abstract}

\section{Introduction}
Measurement-based quantum computing (MBQC) performs universal quantum computation on graph states only by single-qubit measurements, and offers an alternative picture to the quantum information processing usually described in networks of single- and multi-qubit gates \cite{Raussendorf2001,Raussendorf2003}.
This has led to the development of novel approaches in various quantum information theories such as topological error-correction codes \cite{Raussendorf2007,Fujii2015}, blind quantum computing \cite{Broadbent2009a,Morimae2012a}, cross-platform verification \cite{Greganti2021} and quantum circuit depth reduction \cite{Broadbent2009}.
Experimentally, the MBQC paradigm is particularly suited to the optical architectures \cite{Larsen2021, Asavanant2021, Bourassa2021,Takeda2019} with which large-scale cluster states with thousands of modes \cite{Larsen2019,Asavanant2019} has already been demonstrated.
Cold-atom and ion trap quantum computers can also be operated within the framework of MBQC by mid-sequence measurements and feedforward \cite{Graham2022,Bluvstein2022,DeCross2022}.
Furthermore, MBQC was demonstrated on superconducting-qubit quantum computers without feedforward capability \cite{Yang2022}, thus making it a widely relevant paradigm with a variety of potential advantages, e.g.\ significantly smaller execution depth compared to circuits \cite{Broadbent2009}.

In MBQC, an entangled resource state, the graph state, is prepared and the computation is realized by a series of single-qubit measurements only \cite{Raussendorf2003}. 
Any gate network has a corresponding measurement pattern on graph states to realise the same Unitary evolution \cite{Danos2007,Danos2009}, thus making MBQC a universal quantum information processing method.
A remarkable feature of the MBQC is that all Clifford parts of the gate network can be parallelized into a depth of one and that they can be achieved simply by mathematical graph transformations \cite{Raussendorf2003, Anders2006,Aaronson2004}.
This is related to the Gottesman-Knill theorem \cite{Nielsen2010} which states that the quantum information processing task consisting of Pauli basis states, Clifford gates and Pauli measurements are efficiently simulated in classical computers.
To our knowledge, concrete strategies to utilize this feature in MBQC for arbitrary quantum algorithms have not been developed, which is necessary to perform resource-efficient computations.

Here, we extend the measurement calculus \cite{Danos2007,Danos2009} by introducing the \emph{local-Clifford commands}, which we refer to as MBQC on local-Clifford decorated graphs (LC-MBQC).
This is necessary to classically preprocess all Pauli measurements in the measurement pattern with an efficient graph state simulator, since post-measurement states are generally local-Clifford equivalent of graph states.
The resulting local-Clifford decorations only rotate the subsequent non-Pauli measurements, thus maintains the concise operational structure and determinism of the one-way model.

Further, we have implemented a tensor-network (TN) backend for MBQC which is capable of simulating patterns running on thousands of nodes in modest computing resources, without the need for approximation which is often necessary for TN-based circuit simulators.
This is thanks to the concise expression of graph states using matrix product states (MPS) \cite{Vidal2003,Gross2007}, as well as the fact that MBQC completes only with single-qubit operations.

We have assembled the proposed LC-MBQC model, together with several optimization strategies of measurement patterns and their classical simulation backends into an open-source, python-based software library \emph{Graphix} \cite{graphix} which is showcased in Section \ref{sec:overview}. In the following Section \ref{sec:lcmc}, we describe the LC-MBQC and the procedure for classical preprocessing of Pauli measurements in the measurement pattern. Finally, we conclude and provide an outlook in Section \ref{sec:concl}.

\section{Overview}\label{sec:overview}

Here, we first review the MBQC and the measurement calculus in Section \ref{sec:intro} and describe the Graphix library in detail in Sections \ref{sec:patgen} and \ref{sec:patmod}. 
In Section \ref{sec:patsim}, we describe the tensor-network MBQC simulator backend.

\subsection{Introduction}\label{sec:intro}
In Graphix, we describe MBQC with a sequence of commands based on the measurement calculus \cite{Danos2007, Danos2009} with additional local-Clifford command,

\begin{itemize}
	\item node preparation command $N_i$
	\item entanglement command $E_{ij}$
	\item measurement command ${}^t[M_i^{ \lambda, \alpha}]^s$
	\item byproduct correction command $X^s_i, Z^s_i$
	\item local-Clifford command $C_i^{k}$
\end{itemize}
where subscripts $i, j$ are the node (qubit) indices, $\lambda$ specifies the measurement plane (X,Y), (X,Z) or (Y,Z), with the measurement angle $\alpha$, and $s, t$ are the signal domains which are the set of node indices whose measurement results the corresponding operations depend upon \cite{Danos2007}.
In addition to the commands introduced in the measurement calculus, we have added local-Clifford commands $C_i^k$, where $k$ specifies one of 7 Clifford decorations used in the graph state simulator \cite{Anders2006,Elliott2008}, which are necessary to integrate with efficient stabilizer simulator.
As we shall see in Section 3, the Clifford commands only change the measurement angles and do not change the hardware requirements to execute the one-way model, thus none of the advantages of the MBQC are lost.
Any one-way model on a local-Clifford decorated graph state is described by series of the above commands. 
For example, a command sequence realizing Hadamard gate $H$ for an input qubit 1 and output qubit 2 is 
\begin{equation}\label{eq:h}
	\mathcal{M}_H = X_2^1 M_1^{(X,Y),0} E_{12} N_2 N_1,
\end{equation}
which should be read from the right.
For an input state $|\psi\rangle_1$ the operation above is equivalent to the tensor product with qubit 2 in +1 eigenstate of Pauli X operator $|+\rangle_2$, application of CZ-gate, measurement in Pauli X basis and byproduct correction depending on the measurement outcome:
\begin{equation}
	H|\psi\rangle_2 = X_2^{s_1} {}_1\langle \pm_{s_1}| CZ_{12} |+\rangle_2 \otimes |\psi\rangle_1,
\end{equation}
where we denote the Pauli X measurement and subsequent trace out of qubit 1 with ${}_1\langle \pm_{s_1}|$ which depends on the measurement outcome: ${}_1\langle +|$ for $s_1=0$ (measurement outcome $(-1)^{s_1}=+1$) and ${}_1\langle -|$ for $s_1=1$ (measurement outcome of $(-1)^{s_1}=-1$).
The byproduct correction operator $X_2^{s_1}$ is applied conditionally, if the measurement outcome $s_1$ was 1.
For $x$-rotation of a qubit with angle $\xi$, the command sequence is
\begin{equation}\label{eq:xrot}
		Z_3^{s_1} X_3^{s_2} \,\, {}^{s_1}[ M_2^{(X,Y),-\xi} ]\, M_1^{(X,Y),0} E_{23} E_{12} N_3 N_2 N_1.
\end{equation}
In this work, we refer to such a sequence of commands as a \emph{pattern} which is central to the software implementation of the one-way model. 

We further introduce \textit{space} and \textit{depth} of a pattern as follows.
The \textit{space} of the pattern is the number of qubits present during the execution of the pattern. 
The pattern space is first initialized with the number of input qubits, and $N$ commands increment the space while $M$ commands decrement the space, since the measurements are destructive.
The maximum space is the largest space during the execution of a pattern, and is the minimum qubit resource required to run the pattern with qubit reset and reuse \cite{Houshmand2018}.
It is thus often desirable to reduce the maximum space, for example on classical statevector simulators where available memory space bounds the simultaneous qubit count.

The depth of a pattern is the minimum depth (the number of measurement rounds) required to execute the pattern.
For example, Pauli measurements can be performed simultaneously with depth of one \cite{Raussendorf2003}. 
The remaining measurements can be parallelized according to their logical dependence.
The reduction of the depth is beneficial for quantum hardware with limited coherence time.

In the following sections, we describe Graphix in detail, which consists of \emph{generators}, \emph{modifiers} and \emph{simulators} of patterns.

\subsection{Pattern generators} \label{sec:patgen}
At the start of the MBQC programming, we generate the measurement pattern to be performed on the resource graph state, as described below.

\subsubsection{User-specified command sequence}
Any MBQC operations can be expressed by a sequence of commands as described above.
In Graphix, it is possible to manually concatenate commands using \lstinline{Pattern} class, by adding lists specifying each command with \lstinline{add} method. 
Below, we demonstrate the direct programming of pattern Eq.\ \eqref{eq:h}.

\begin{lstlisting}[language=Python, caption=User-defined pattern., label=arbitrary_pattern]
from graphix import Pattern
# Measurement pattern for Hadamard gate
pattern = Pattern()
pattern.add(['N', 0])
pattern.add(['N', 1])
pattern.add(['E', (0, 1)])
pattern.add(['M', 0, 'XY' ,0.,  [], []])
pattern.add(['X', 1 ,[0]])
\end{lstlisting}

\subsubsection{Transpile from a gate network}\label{sec:transpile}
With \lstinline{graphix.Circuit} class, one can create a gate sequence and transpile into a measurement pattern, with pre-configured byproduct sequences that ensure determinism \cite{Raussendorf2003,Danos2009}, using \lstinline{transpile} method:
 
\begin{lstlisting}[language=Python, caption=Transpiling a pattern from a gate network., label=arbitrary_pattern]
from graphix import Circuit
import numpy as np
circuit = Circuit(2)
circuit.cnot(0,1)
circuit.rx(0, np.pi/4)
circuit.cnot(1,0)
pattern = circuit.transpile()
\end{lstlisting}

\subsubsection{Pattern generation with flow or gflow}
Designing the byproduct sequence and ensuring the determinism is the most critical aspect of MBQC programming, however, it is a time-consuming process. 
As such, we implement an algorithm to find \textit{flow} \cite{Danos2006} or \textit{generalized flow} (\textit{gflow}) \cite{Browne2007} on open graph \cite{Mhalla2008} in the \lstinline{graphix.gflow} module, with which one can automatically generate the command sequence only from the measurement angles and shape of the graph, i.e. this algorithm automatically generates the byproduct correction sequence that ensures determinism. 
This can be programmed as follows, where this example generates the pattern to realize the CNOT gate followed by the Hadamard gate applied to the target qubit:

\begin{lstlisting}[language=Python, caption=flow- or gflow-based pattern generation., label=arbitrary_pattern]
from graphix import generate_from_graph
import networkx as nx
g = nx.Graph()
g.add_nodes_from([0,1,2,3,4])
g.add_edges_from([(0,2),(1,2),(1,3),(2,4)])
angles = {0: 0., 1: 0., 2: 0.}
pattern = generate_from_graph(g, angles, inputs=[0,1], outputs=[4,3])
\end{lstlisting}

\subsection{Pattern modifiers}\label{sec:patmod}
Pattern modifiers manipulate the command sequences in the \lstinline{Pattern} object, and optimize their properties such as command length, space and depth.

\subsubsection{Standardization and signal shifting}
The \emph{standard} pattern is the one where the commands are sorted in the order of $N$, $E$, $M$, followed by the byproduct and Clifford commands $X, Z, C$ which can be in arbitrary order. 
One can \emph{standardize} any command sequence by the commutation relations of individual commands, as described in \cite{Danos2007,Danos2009}.
Furthermore, the \emph{signal shifting} procedure reduces the interdependence of measurement commands \cite{Danos2007}.
These can be performed by \lstinline{standardize()} and \lstinline{shift_signals()} methods of \lstinline{Pattern} class and the two operations scale with the number of total commands $N_c$ as $\mathcal{O}(N_c^5)$ \cite{Danos2007}.

When translating a gate network to a pattern, we can utilize the equivalence to unitary gate sequence to simplify the standardization procedure. 
This results in a standardized pattern with no $t$-domain, the same as the one following the signal shifting.
The computation of this process scales much preferably with $\mathcal{O}(N_g^3)$, where the number of gates $N_g$ is typically much smaller than $N_c$.
We have implemented this in \lstinline{Circuit} class as \lstinline{standardize_and_transpile()} method.
	
\subsubsection{Pattern optimization}
We provide further pattern optimization procedures.
Firstly, Pauli measurements of a pattern can be classically preprocessed without simulating the full quantum state thanks to efficient graph state simulators, and is thus implemented as a pattern modifier.
This is called by \lstinline{Pattern.perform_pauli_measurements()} after which the length of the pattern sequence is significantly reduced while kept deterministic.
We describe the detailed internal procedure in Section \ref{sec:lcmc}. 

We also provide a method for the minimization of pattern space, by the reordering of $N, E$ and $M$ commands.
Essentially, this is achieved by delaying the preparation of qubits until their immediate neighbors are measured \cite{Houshmand2018}.
We reorder measurement commands according to their feedforward dependence structure to further minimize the maximum space.
In Graphix, this can be performed by \lstinline{Pattern.reduce_space()}.

Another optimization method we provide is the parallelization of commands for reduced execution depth.
This is obtained by performing all logically independent measurements at once, and can be called with \lstinline{Pattern.parallelize_commands()}.

\subsection{Pattern simulator}\label{sec:patsim}
The \lstinline{graphix.PatternSimulator} class executes the pattern using classical simulation backends; we have implemented statevector and matrix product state (MPS) backends (see Sec.\ \ref{sec:mps}).
These can also be called from pattern class by \lstinline{pattern.simulate_pattern()}.

Further, it is straightforward to add quantum hardware and hardware emulators as pattern execution backends.
This is because the pattern sequence can be run only with standard operations such as qubit preparation, CZ-gate and single-qubit measurements (and that byproduct and Clifford commands can be straightforwardly replaced by the rotation of final readout measurement).

\subsubsection{Tensor network backend}\label{sec:mps}
It has been known that certain types of quantum states have an efficient expression by tensor networks (TN) such as matrix product states (MPS) \cite{Vidal2003,Schollwock2011, Shi2006, Markov2008, Nest2007, Morimae2012}.
This allows fast simulation of quantum circuits with a limited amount of entanglement \cite{Vidal2003} and has led to a recent development of TN-based quantum circuit simulators \cite{Fang2022, Vincent2022, roberts2019, Liu2021}.
However, they are often limited in their applicability since highly entangled states can no longer be efficiently expressed using MPS without approximation \cite{Schollwock2011,Zhou2020}.

Such a limitation of the circuit MPS simulation could be mitigated by translating into MBQC since any quantum algorithm (circuits) has corresponding measurement patterns running on graph states with only two-body entanglements and computation completes by single-qubit operations.
This allows for efficient simulation with potentially much wider applicability than standard TN-based circuit simulations.
Specifically, MBQC on tree graphs \cite{Shi2006} and graph states with limited \textit{Schmidt-rank width} \cite{Nest2007}, were known to be efficiently simulatable using MPS. 
However, software libraries for demonstrating and benchmarking this idea is not widely available.

In Graphix, we implement an efficient preparation method of graph states on MPS \cite{Gross2007, Morimae2012} to realize a significant speedup of MBQC pattern simulation compared to standard state vector simulator. 
Further, pattern optimization can be tailored for MPS simulator characteristics, e.g. by limiting the space of the pattern.
As such, Graphix serves as a comprehensive platform to test the boundary of applicability for the MBQC-MPS simulation method.
An intriguing advantage in comparison to the circuit-based TN technique is that the bond dimension \cite{Schollwock2011} of two is sufficient to exactly describe the resource state, as well as that only single-qubit (single-tensor) operations are performed after graph state preparation for which MPS runs efficiently.
With our current implementation, MBQC on graphs with thousands of nodes can be simulated with modest computing resources, on which only up to $\sim$20 qubits can be prepared at once as a statevector.
Further, we are developing an optimized MBQC-MPS simulator to run on a GPU cluster, all of which will be described and benchmarked in detail in our separate work  \cite{Fukushima2022}.

\subsection{Relation to previous works}\label{sec:prev}

The possibility of eliminating Pauli measurements in one-way quantum computing was pointed out in the initial work by Raussendorf \cite{Raussendorf2003} based on the correspondence to stabilizer code, however, the concrete strategy to integrate this idea into the computational model of MBQC has not been developed so far, to our knowledge.
Recent work by Houshmand et al \cite{Houshmand2018} showed only the minimum number of qubits required in the most general case and suggested the presence of determinism using the notion of gflow \cite{Browne2007} for a single example.
Another recent work by Ferguson et. al \cite{Ferguson2021} optimize only a few variational algorithms, without devising the methodology for general quantum algorithms.

Other simulators for the measurement-based model, Paddle Quantum \cite{paddle} and MCBeth \cite{Aidan2022} were recently developed to implement the measurement calculus.
The Graphix package offers a significantly advanced capability by incorporating the stabilizer simulator with LC command, novel pattern optimization methods and tensor-network backend.

We note that the measurement calculus also naturally describes other measurement-based models \cite{Danos2009} such as the teleportation-based model \cite{Nielsen2003}. Thus Graphix will flexibly accommodate a wide range of measurement-based architectures.

\begin{figure*}[t]
	\centering
	\includegraphics[width=0.9	\textwidth]{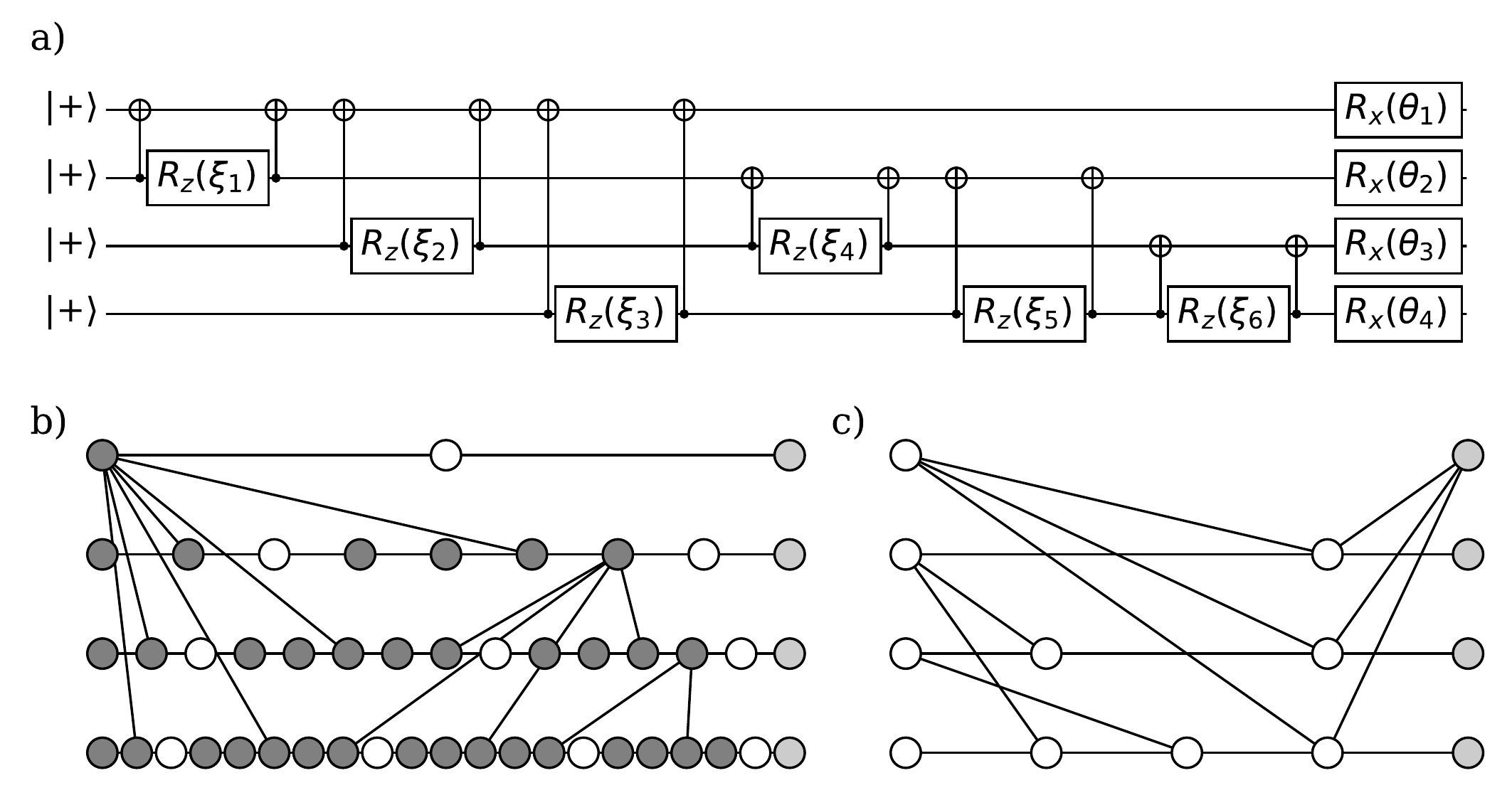}
	\caption{\label{fig:qaoa} The quantum approximate optimization algorithm (QAOA) circuit and the resource state for MBQC implementation before and after the graph transformation. 
		a) One layer of QAOA circuit for a fully-connected graph with four vertices.
		b) The graph state to perform the QAOA with the one-way model. Four leftmost nodes (qubits) are the input and four rightmost nodes are the output qubits. The nodes to be measured in Pauli bases are indicated by black circles while the ones with non-Pauli measurements are in white. Output qubits are in gray.
		c) The graph state required to execute the pattern generated using \lstinline{Pattern.perform_pauli_measurements()}, following the procedure \ref{sec:node_remove}. 
		This is the minimal resource state that performs the same deterministic quantum computation as the graph state shown in b).
		The number of nodes reduces from 48 to 14 and the minimum depth of measurement rounds, obtained by \lstinline{pattern.parallelize_commands()}, reduces from 3 to 2 as a result of the node elimination procedure.}
\end{figure*}	

There are numerous stabilizer simulators that perform fast Clifford operations to stabilizer states with a large number of qubits, such as graph-state \cite{Anders2006} and stim \cite{Gidney2021}. 
We have implemented in Graphix the decorated graph state simulation proposed in \cite{Elliott2008,Elliott2009} to efficiently perform the Pauli measurements of graph states.
This implementation also allows us to toggle through the set of equivalent graphs, a set of different local-Clifford-decorated graph states representing exactly the same stabilizer state, and this procedure can be used to minimize the connectivities of the graph and thus optimize subsequent MBQC operations.

ZX-calculus \cite{Coecke2011} is a graphical tool for circuit optimizations, which also has application in optimizing MBQC \cite{McElvanney2022} while preserving the existence of the flow.
While our method of Pauli measurement elimination in Sec. \ref{sec:node_remove} does not necessarily preserve the flow, it does preserve the determinism by considering the measurements on the graph state simulator as part of the pattern execution and this reduces a much larger number of measurement commands (i.e. the size of the graph state).
For users to be able to evaluate ZX-calculus for MBQC pattern sequence generated with Graphix, we provide a method to export from a measurement pattern to openQASM \cite{Cross2017}, which can be loaded into python library PyZX \cite{kissinger2020Pyzx}.

\section{Measurement calculus with local-Clifford command $C_i^k$}\label{sec:lcmc}

In this section, we introduce the measurement calculus \cite{Danos2009} with local Clifford commands, which is essential to incorporate the fast graph state simulator into the one-way model to preprocess all Pauli measurements.

\subsection{Classical preprocessing of Pauli measurements}\label{sec:node_remove}
To perform Pauli measurements of a pattern using a graph state simulator, and to obtain a pattern with significantly reduced command length, the following procedure is used:
\begin{itemize}
	\item[1.] Perform standardization of the pattern. 
	\item[2.] Perform signal shifting to eliminate the dependence of Pauli measurements on non-Pauli measurements, and bring Pauli measurements to the front (just after entanglement commands).
	\item[3.] Extract the $N$ and $E$ commands of the pattern to initialize a graph state simulator. 
	Perform Pauli measurements on the graph state simulator and remove corresponding commands from the measurement pattern. 
	Choose the measurement results $s_i=0$ where possible (except for isolated nodes with 100$\%$ chance of being measured in $|-\rangle$ state of the corresponding measurement basis).
	\item[4.] From the resulting local-Clifford decorated graph state in the graph state simulator, insert $N$, $E$ and $C$ commands to the front of the pattern.
\end{itemize}

The resulting command sequence is runnable and deterministic, since the measurement results are known and can be used for subsequent feedforward operations where necessary.
As we describe below, the local-Clifford commands $C$ only rotate the subsequent measurements, and there is no need to prepare local-Clifford decorated graphs quantum mechanically; the pattern runs exactly the same way as the standard measurement calculus, with rotated measurement angles.

\subsection{Local-Clifford command}
Here, we describe the effect of local-Clifford decorations (commands) on the subsequent MBQC operations (rotation of measurement angles, including that of final readout measurements).

Single-qubit Clifford operators $O_C$ map a Pauli operator $O_p \in \{X,Y,Z\}$ to an element in $\mathcal{P} =\{\pm\}\times \{X,Y,Z\}$ under conjugation with $O_C$, i.e.
\begin{equation}
O_C O_p O_C^{\dagger} \in \mathcal{P}.
\end{equation}
For a single-qubit projection operator $P_{\bm{r}} = (I + r_1 X + r_2 Y + r_3 Z)/2$ in an arbitrary projection angle specified by a vector with unit length $\bm{r} = (r_1, r_2, r_3)$, the following holds:
\begin{equation}
O_C^\dagger P_{\bm{r}} O_C = P_{\bm{r}'},
\end{equation}
with a rotated vector $\bm{r}'$ (see Appendix 1 for the list of transformation rules for each $O_c$ used in Graphix).
Thus, the local Clifford commands rotate the measurement angles and the following rules apply to the local-Clifford commands:
\begin{eqnarray}\label{eq:lc_command}
C_i^k [M_i^{\lambda,\alpha}] &=& [M_i^{\lambda',\alpha'}], \nonumber \\
C_i^k [M_j^{\lambda,\alpha}] &=& [M_j^{\lambda,\alpha}] C_i^k, \ i\neq j
\end{eqnarray}
where the relationships between $(\lambda, \alpha)$ and $(\lambda', \alpha')$ follow from Table\,\ref{tab:clifford} in Appendix \ref{sec:appA}.
If the local Clifford command operates on the output nodes, it is applied to the final output states or rotate the final readout measurement basis.

\subsection{Example: Variational quantum algorithm}
We demonstrate the procedure \ref{sec:node_remove} with quantum approximate optimization algorithm (QAOA), which is a heuristic variational algorithm for optimization problems such as max-cut \cite{Farhi2014}.
We show in Fig.\ref{fig:qaoa} a) the typical circuit to perform an evaluation of variational parameters $\bm{\xi}$ and $\bm{\theta}$ for the max-cut problem of a fully-connected graph with four nodes.
We transpile this circuit into a measurement pattern using \lstinline{graphix.Circuit} class as described in Section \ref{sec:transpile}, which results in the command sequence that requires the total graph state shown in Fig.\,\ref{fig:qaoa} b).
Using the procedure in Section, \ref{sec:node_remove}, the length of the command sequence reduces from 156 to 50, which requires a much smaller graph state shown in Fig.\,\ref{fig:qaoa} c).
This can be performed by the following python code:

\begin{lstlisting}[language=Python, caption=Optimizing single-layer QAOA pattern., label=qaoa]
from graphix import Circuit
import networkx as nx
import numpy as np
n = 4
xi = np.random.rand(6)
theta = np.random.rand(4)
g = nx.complete_graph(n)

circuit = Circuit(n)
for i, (u,v) in enumerate(g.edges):
	circuit.cnot(u, v)
	circuit.rz(v, xi[i])
	circuit.cnot(u, v)
for v in g.nodes:
	circuit.rx(v, theta[v])

pattern = circuit.transpile()
pattern.standardize()
pattern.shift_signals()
pattern.perform_pauli_measurements()
out_state = pattern.simulate_pattern()
\end{lstlisting}

\begin{figure}[t]
	\centering
	\includegraphics[width=0.99	\linewidth]{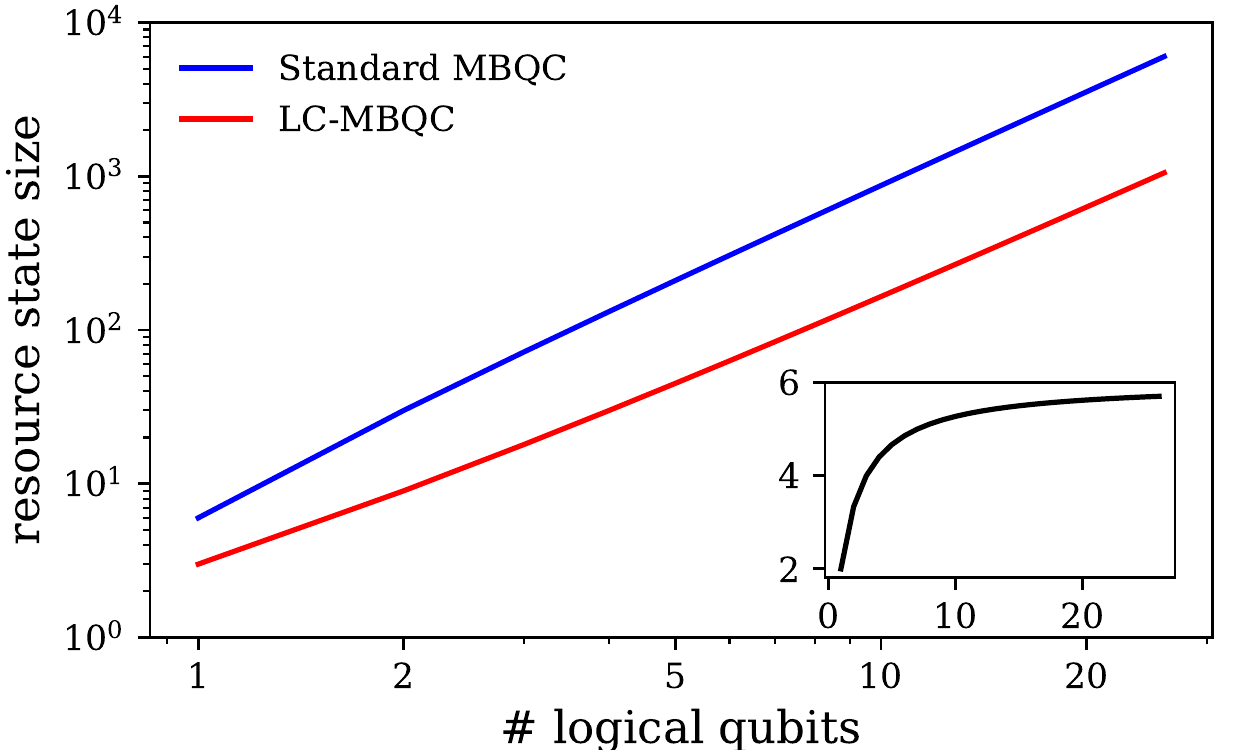}
	\caption{\label{fig:scaling} The number of nodes in the resource state to perform the three-layer QAOA circuit for the max-cut problem of fully-connected graphs, before (Standard MBQC) and after (LC-MBQC) the node elimination procedure shown in Sec.\ref{sec:node_remove}. 
		The inset shows the ratio of the graph state size before and after the procedure.}
\end{figure}

In Fig.\ref{fig:scaling}, we plot the number of nodes in the resource state before and after \lstinline{pattern.perform_pauli_measurements()}, translated from QAOA circuits for max-cut problems of complete graphs with different sizes (number of logical qubits). 
The resource state size required to perform MBQC is significantly reduced by up to a factor of 6.
Generally, this method preprocesses all Pauli measurement corresponding to Clifford gates in the original gate network, and thus its effect is more significant for quantum algorithms involving a large number of Clifford gates.

\section{Conclusion and Outlook}\label{sec:concl}
We have introduced LC-MBQC, a measurement calculus formulation of MBQC with local Clifford commands. 
With this, we have naturally integrated an efficient graph state simulator as an MBQC pattern sequence preprocessing routine which allows the significant reduction of the size of the resource state required to execute universal quantum computations by single-qubit measurements only.
We have implemented further optimization strategies for pattern sequence, such as faster standardization and algorithms to reduce the pattern space or the execution depth.
We further provide an MPS simulator backend, which improves the classical simulation of MBQC.

A natural extension of Graphix is the implementation of noise models for specific hardware implementations such as CV photonic architectures with Bosonic codes.
With this, we plan to integrate error correction algorithms for the one-way model, such as topological error correction \cite{Raussendorf2007}, into the measurement pattern architecture.
Furthermore, it is possible to express and simulate unique approaches in the one-way model such as blind quantum computations \cite{Broadbent2009,Morimae2012} and distributed MBQC \cite{Danos2007,Hondt2009} thanks to the flexibility of the pattern-based construction, thus making Graphix a suitable platform for further research into the measurement-based model.

\section*{Acknowledgement}
S.S. is supported by EPSRC Grant Reference EP/S013105/1.
Authors thank Christopher Foot for reading the manuscript, Suguru Endo, Takuji Hiraoka and Yoshiki Matsuda for insightful discussions and Ryosuke Suda for contribution to the library.

\bibliographystyle{apsrev4-2}
\appendix

\section{Clifford commands}\label{sec:appA}

Pauli measurements of a graph state performed in the graph state simulator \cite{Elliott2008} result in a graph state with local-Clifford decorations up to one $H$, $S$ and $Z$ gates on each nodes i.e. the post-measurement state is \cite{Elliott2009}
\begin{equation}
|g\rangle = \prod_{i\in\mathcal{H}_g} H_i \prod_{j\in\mathcal{S}_g} S_j \prod_{k\in\mathcal{Z}_g} Z_k \prod_{(l,m)\in\mathcal{E}_g} CZ_{l,m} \,\, |+\rangle^{\otimes n},
\end{equation}
where $\mathcal{H}_g, \mathcal{S}_g$ and $\mathcal{Z}_g$ are the sets of nodes with decorations with each Clifford gates, $\mathcal{E}_g$ is the set of edges and $n$ is the number of nodes in the graph.
There are 7 unique Clifford commands to be used in the procedure described in Section\,\ref{sec:node_remove} for which we list their effect on Pauli $X, Y$ and $Z$ gates as follows.
The rotation of measurement angle \eqref{eq:lc_command} follows from Table \ref{tab:clifford} for any $k$, $\lambda$ and $\alpha$.

\begin{table}[h]
	\begin{center}
		\begin{tabular}{lcccc}
			\toprule{}
			cmd & $O_C$ & $O_C^\dagger X O_C$ & $O_C^\dagger Y O_C$ & $O_C^\dagger Z O_C$ \\[-1em]
			\colrule{}
			$C^0$ & $H$ & $Z$ & $-Y$ & $X$ \\
			$C^1$ & $S$ & $-Y$ & $X$ & $Z$  \\
			$C^2$ & $Z$ & $-X$ & $-Y$ & $Z$ \\
			$C^3$ & $HS$ & $Z$ & $-X$ & $-Y$  \\
			$C^4$ & $HZ$ & $Z$ & $Y$ & $-X$  \\
			$C^5$ & $SZ$ & $Y$ & $-X$ & $Z$  \\
			$C^6$ & $HSZ$ & $Z$ & $X$ & $Y$ 
			\botrule{}
		\end{tabular}
	\end{center}
	\caption{\label{tab:clifford} Effect of single-qubit Clifford operators on Pauli operators.}
\end{table}

\end{document}